\documentclass[12pt]{article}

\PassOptionsToPackage{numbers, compress}{natbib}
\usepackage[preprint]{neurips_2020}

\usepackage{booktabs}       
\usepackage{amsfonts}       
\usepackage{nicefrac}       
\usepackage{graphicx}
\usepackage{xcolor}
\usepackage{enumitem}
\usepackage{mathtools}

\DeclarePairedDelimiter\floor{\lfloor}{\rfloor}
\usepackage{wrapfig}
\usepackage{url,lineno,microtype,subcaption}
\usepackage[onehalfspacing]{setspace}
\usepackage{blindtext}
\usepackage[ruled,vlined]{algorithm2e}

\usepackage{cite}
\usepackage{bm}

\usepackage{amsmath}

\makeatletter
 \def\@textbottom{\vskip \z@ \@plus 1pt}
 \let\@texttop\relax
\makeatother

\title{Neuro-PC: Causal Functional Connectivity from Neural Dynamics}
\author{
  Rahul Biswas\\ 
  Department of Statistics\\
  University of Washington\\
  Seattle, WA, 98195\\
  \texttt{rbiswas1@uw.edu}\\
  \And
  Eli Shlizerman \\ 
  Department of Applied Mathematics\\
  Department of Electrical \& Computer Engineering\\  University of Washington\\
  Seattle, WA, 98195 \\
  \texttt{shlizee@uw.edu} \\
}

\begin{document}

\maketitle

\begin{abstract}
Functional connectome extends the anatomical connectome by capturing the relations between neurons according to their activity and interactions. When these relations are causal, the functional connectome maps how neural activity flows within neural circuits and provides the possibility for inference of functional neural pathways, such as sensory-motor-behavioral pathways. While there exist various information approaches for non-causal estimations of the functional connectome, approaches that characterize the causal functional connectivity - the causal relationships between neuronal time series, are scarce. In this work, we develop the Neuro-PC algorithm which is a novel methodology for inferring the causal functional connectivity between neurons from multi-dimensional time series, such as neuronal recordings. The core of our methodology relies on a novel adaptation of the PC algorithm, a state-of-the-art method for statistical causal inference, to the multi-dimensional time-series of neural dynamics. We validate the performance of the method on network motifs with various interactions between their neurons simulated using continuous-time artificial network of neurons. We then consider the application of the method to obtain the causal functional connectome for recent multi-array electrophysiological recordings from the mouse visual cortex in the presence of different stimuli. We show how features of the mapping can be used for quantification of the similarities between neural responses subject to different stimuli.

\end{abstract}

\section{Introduction}
The anatomical connectome is the physical map of all the neurons and the connections between them within a particular part of the brain~\citep{bargmann2013connectome}. For example, the connectome of \textit{Caenorhabditis elegans} somatic nervous system contains all the somatic neurons (279) and enumerates the synaptic connections between them, such as glutamatergic, cholinergic, GABA, and also gap junctions~\citep{white1986structure,varshney2011structural,cook2019whole}. The mapping of the anatomical connectome can be achieved with the help of imaging techniques and computer vision technologies at different scales~\citep{shi2017connectome, sarwar2020towards,xu2020connectome}. 
While the anatomical connectome includes the backbone information on the possible ways how neurons could interact, it does not fully reflect the ``wiring diagram of the brain", which is expected to incorporate the dynamic nature of neurons' activity and their interactions~\citep{lee2011specificity,kopell2014beyond,kim2017neural,kim2018movingcelegans}.

It is therefore desirable to obtain a more complete map reflecting the particular functions that neurons perform. Such a mapping is the functional connectome (FC) and it represents the network of associations between the neurons with respect to their activity over time \citep{reid2012functional}. FC is expected to include and facilitate inference of the governing neuronal pathways essential for brain functioning and behavior~\citep{finn2015functional}. Two neurons are said to be functionally connected if there is a significant relationship between their activity over time, where the activity can be observed by recording from neurons over time and measured with various measures~\citep{shlizerman2012neural}. In contrast to the anatomical connectome, the functional connectome cannot be directly observed or mapped, since the transformation from the activity to a network of associations is non-unique. This indicates that even if the basic physical attributes of a neurobiological system have been extensively characterized, the functional units and associations between them could remain poorly understood. To address these challenges, there has been an increasing interest in the inference of brain network interactions described in the functional connectome at different levels of neural organization \citep{reid2019advancing, foti2019statistical, zuo2012network}.

Existing methods to characterize FC include approaches based on measuring correlations, such as pairwise correlation~\citep{rogers2007assessing,preti2017dynamic}, or sparse covariance matrix that is comparatively better than correlations given limited time points~\citep{xu2015dynamic,wee2016diagnosis}.
Furthermore, for such scenarios, regularized precision matrix approaches were proposed to better incorporate conditional dependencies between the time courses ~\citep{varoquaux2010brain,smith2011network}, where the precision matrix is inferred by a penalized maximum likelihood \citep{friedman2008sparse} to promote sparsity. For example, Allen et. al \citep{allen2014tracking} proposed to find the FC by regularized precision matrix between the time-courses after Independent Component Analysis to account for time-varying noise. These methods were applied to functional Magnetic Resonance Imaging (fMRI) data to obtain the functional connectivity between different broad regions or voxels of the brain. In addition, the aforementioned methods have been extended to obtain FC over short windows of time and to compile them to give a dynamic FC which changes over time \citep{barttfeld2015signature,cribben2012dynamic,marusak2017dynamic,rashid2014dynamic,wee2016sparse,cribben2013detecting,damaraju2014dynamic}. Furthermore, consideration of dominant neural patterns has been proposed to replace the consideration of correlations to measure association between neurons. In particular, it was proposed to obtain a low-dimensional embedding through application of SVD on neural activity and to translate the patterns to conditional probabilities for the construction a probabilistic graphical model that corresponds to the FC~\citep{liu2017functional}.

The prevalent research on FC, outlined above, deals with finding associations between neural signals in a non-causal manner. That is, we would know that a neuron A and a neuron B are active in a correlated manner, however, we would not know the answer to whether the activity in neuron $A$ causes neuron $B$ to be active ($A\rightarrow B$), or is it the other way around ($B\rightarrow A$)? Or, is there a neuron $C$ which intermediates the correlation between $A$ and $B$ ($B\leftarrow C\rightarrow A$)? These three questions distinguish causation from correlation. In the current paper, we aim to incorporate the aspect of causality by finding the \emph{causal functional connectome}, which would specifically answer the aforementioned causal questions. At the outset, insights from existing FC approaches \citep{valdes2011effective,ramsey2010six} communicate that the phenomenon of theoretical interest for FC research is ultimately to find the causal connection between the neural entities. It is noteworthy that results of association have also been used to make causal implications, for example, correlated brain regions have been considered as causal entities \citep{yeo2011organization,power2011functional, power2013control, smith2009correspondence}. 
This suggests that it would be helpful to make causal reasoning explicit in the FC methodology.

\section{Related Work}
In the literature of FC research, there have been contributions to infer causality using approaches like Granger Causality \citep{ tank2017granger,seth2015granger,friston2013analysing,xu2016learning}, and Dynamic Causal Modeling \citep{sharaev2016effective}. Dynamic Causal Modeling treats the brain as a deterministic non-linear dynamical system, incorporating interactions due to the experimental equipment, and determines the strength of causal effect along the anatomical connections and changes due to external perturbations, however it requires a detailed mechanistic description of neuronal dynamics~\citep{stephan2010ten}. On the other hand, Granger Causality is a widely used statistical method aiming to infer causality, and recent extensions to Granger Causality have also been proposed for non-linear neural time series scenario by levying recurrent neural networks \citep{tank2018neural}. However between a pair of time series Granger Causality gauges whether one series is helpful in predicting the other and the former has temporal precedence \citep{stokes2017study,dahlhaus2003causality,guo2018survey}, and there are studies expressing concern on the extent it \textit{causes} the data in the other \citep{friston2009causal}. In several causal situations, it fails to give accurate result, for example, in the presence of a common cause: Hot days cause more people to swim, and hot days cause more heat strokes, so for the series on number of people swimming and the number of heat-strokes, the former would be predictive of the later and so swimming would be Granger Causal for heat-stroke, while they are not actually causal~\citep{maziarz2015review}. Notwithstanding the limitations, Granger Causality is a well-known method in the neural time series scenario~\citep{guo2020granger,qiao2017functional}. In this work we explore modeling causality by another approach, known as the Probabilistic Graphical Models \citep{koller2009probabilistic}, which gives a nuanced alternative formulation, yet lacking in  research and application in the neural dynamics scenario and is recently gathering interest \citep{liu2018functional, iyer2013inferring}, although finds wide application in other fields like genetics and genomics \citep{wang2017potential,sinoquet2014probabilistic, mourad2012probabilistic, wang2005new, friedman2004inferring}.  

\begin{figure}[t]
    \centering
    \includegraphics[width=0.8\textwidth]{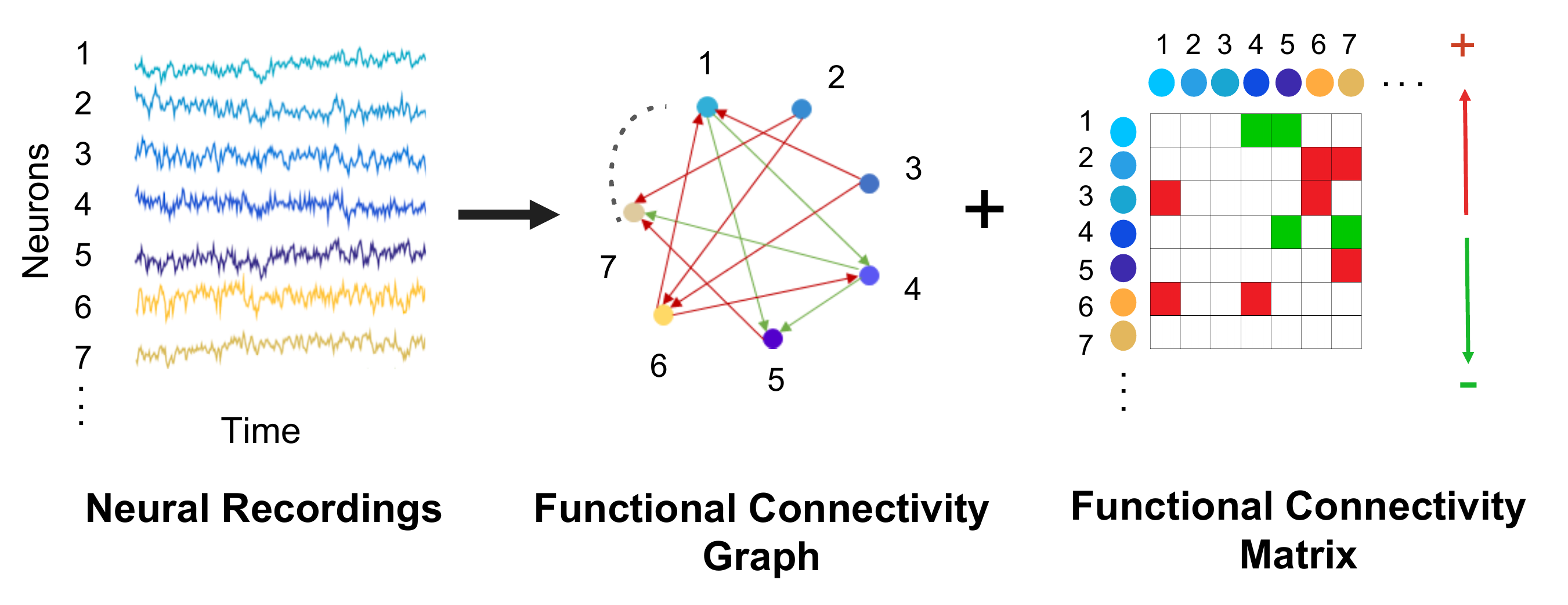}
    \caption{Causal FC inference receives input of neural recordings, e.g., peri-stimulus time histograms (firing-rates), over time and outputs the Causal FC in terms of the \textit{FC Graph} and \textit{FC Matrix}. The \textit{FC Graph} represents the causal connections between the neurons, with nodes as neurons and directed edges between the nodes as the direction of causal connectivity between them. For example, the directed edge starting from neuron 3 to neuron 1, in this figure, represents that activity of neuron 3 has a causal effect on the activity of neuron 1. The \textit{FC matrix} quantifies the strength of the causal effect. For example, a value of 0 in cell (0,1) represents that there is no directed edge from 0 to 1, and a non-zero value in cell (3,1) represents that there is a directed edge from 3 to 1 and the non-zero value is the strength of that directed edge.}
    \label{fig:fig1}
\end{figure}

In the statistics literature, Spirtes-Glymour-Scheines (SGS) algorithm \citep{glymour1991causal}, Peter-Clark (PC) algorithm \citep{spirtes2000causation}, Greedy Equivalence Search (GES) algorithm \citep{chickering2002optimal} are popular to infer causality by outputting a directed acyclic graph among the variables of interest. Under the Causal Markov Condition, which says that - a variable X is independent of every other variable (except X's effects) conditional on all of its direct causes - it follows that all the common causes among the variables, as illustrated in the preceding paragraph, will be included in the directed acyclic graph outputted by these algorithms. Furthermore, any intervention into one variable would change only those variables who are causally ``downstream" to the intervened variable \citep{spirtes2000causation}. However, these algorithms are developed for static data and are not suitable to be directly applied to infer causal relationships in data with dependency in the temporal domain, such as multidimensional time-series. Aspects like non-stationarity of the time series would allow the output of these constraint-based causal inference algorithms to output the true causal structure. This creates a gap in the literature of algorithmic approaches to infer causal FC from multi-neuron recordings over time.

In this paper, we develop a novel methodology for causal inference in the multi-dimensional time-series scenario, coined Neuro-PC, and show applications to neural dynamics (See Figure \ref{fig:fig1}). We propose to use the PC algorithm \citep{spirtes2000causation} as a starting point for building the methodology, given it is one of the widely used causal inference algorithms on static data, it is consistent under i.i.d. sampling with no latent confounders, it is faster than the SGS algorithm and it is not greedy or structurally restrictive like GES \citep{kalisch2007estimating}.
Neuro-PC relies on adapting the PC algorithm to the multi-dimensional neural time-series setup, by following processes such as time-delay, bootstrapping and neuron ablation, to make it robust and well suited to the setup. We validate the proposed methodology on simulated neural signals in form of firing-rates from continuous time artificial neural networks set according to particular motifs. We show the importance of causality in being able to recover the generating motifs and that
Neuro-PC is able to recover the causal relationships and the generating motifs comparatively well. We further use the method to obtain the causal FC among sampled neurons in mice from electrophysiological neural signals.

We organize the paper as follows. The following section gives a brief overview of causal inference for static variables using the PC algorithm. Section 3 discussed the Neuro-PC algorithm. Section 4 outlines the results of application on simulated data and validation of performance of the algorithm, followed by Section 5 with application to neurobiological data. We conclude with a discussion.



\begin{figure}[t]
    \centering
    \includegraphics[width =0.95\textwidth]{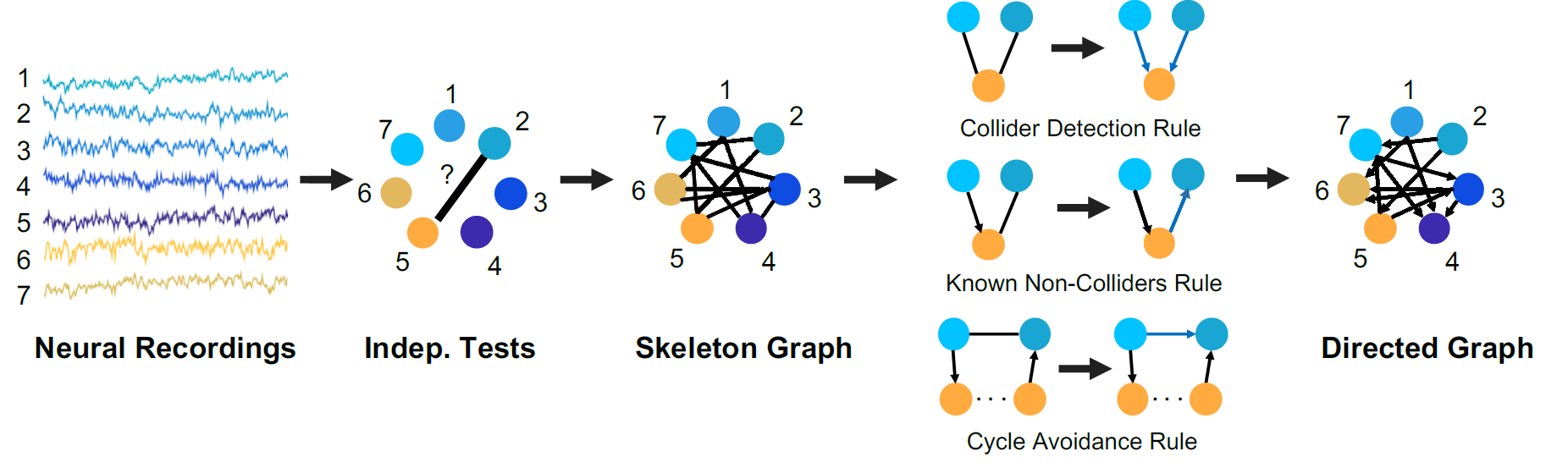}
    \caption{Steps of the PC algorithm which infers the directed FC graph from raw data.
    }
    \label{fig:fig2}
\end{figure}
\section{Causal Functional Connectivity Inference for Static Variables - The PC Algorithm}
In this section, we provide a concise summary of the PC algorithm that will be used in further sections to develop our proposed method. The PC algorithm \citep{spirtes2000causation}, outlined in Figure \ref{fig:fig2}, is a popular statistical method to infer causal connections between static random variables. The PC algorithm is a constraint-based approach, in which any consistent statistical test for conditional independence is used to select the connectivity graph among variables of interest. For Gaussian random variables, the standard choice for such a test is the Fisher's Z-transform, as constructed in \citep{kalisch2007estimating}. 

For our purpose of extending the PC algorithm for multi-neuron recordings, we describe the PC algorithm in the context of neural firing rates and depict it in Figure~\ref{fig:fig2}. The first step is to determine the undirected edges between pairs of neurons. We start with an empty graph and decide whether to put an undirected edge between a particular pair of neurons, say neurons 2 and 5 in Figure ~\ref{fig:fig2}. In order to determine whether such edge should exist between the two neurons, we perform a series of hypothesis tests. We use conditional independence test by Fisher's Z-score. We first test if the time series for neurons 2 and 5 are independent. If they are not, we put an edge between the two. If they are independent, we test with each other single neuron, if they are conditionally independent given that neuron. Say for the pair of neurons 2 and 5, we test conditional independence given neuron 1, if still independent, given neuron 3, and so on. If so far conditional independence holds, we test if they are conditionally independent given pairs of neurons, followed by triples, and so on. If the conditional independence test fails at any step, we connect the neurons 2 and 5 by an edge. We repeat the process for each pair of neurons. The obtained undirected graph is called the skeleton graph. Using rules such as the Collider Detection rule, Known Non-Colliders rule and Cycle Avoidance rule, the skeleton graph is converted to assign directed connectivity to the graph edges.

While the PC algorithm is simple to implement and efficient variants exist, it finds poor relevance in the context of neural time series since there are inter-temporal causal dependencies within and between the time-series, and there are no replications for the recordings at each time that make the results relying on a single instance and hence unreliable. In our proposed method we consider both these aspects of making the PC algorithm a suitable approach for application to the multi-dimensional neural time series data.

\section{The Neuro-PC algorithm}
Neuron's activity at the previous time point $t-1$ affects recordings at next time point $t$ and possibly further time points of the same neuron and of other neurons. To incorporate these effects we would like to keep different variables separately representing previous time point of neuron i, following time point of neuron i and following time point of neuron j, in our setup. With this intuition, we disassociate each neuron signal into time delayed signals, which are distinct signals whose sampled time points have a fixed time delay between the signals. Denoting the original signals by $S_{ti}$ for time point $t$ for neuron $i$, where $1\leq i\leq N, 0\leq t\leq T-1$, $N$ being the  number of neurons, and $T$ the number of time points recorded for each neuron, the time delayed signals are given by
\begin{equation*}
  S_{ti}^\tau = S_{kt +\tau,i},
\end{equation*} 
where $S_{ti}^\tau$ is the $\tau$-th time delayed signal corresponding to neuron $i$, $1\leq \tau \leq k$ and $k$ is the maximum time delay incorporated. $k$ represents the maximum delay of interactions among neurons and is predetermined. Instead of the original time series for the neurons $S_i$, we will consider the disassociated time delayed series $S_{ti}^\tau$ for each neuron in the next steps. In this way a dependency of the $\tau+1$-th delayed variable for neuron $i$, taking values $\{S_{ti}^{\tau+1}:1\leq t\leq T-1\}$ non-sequentially, on the $\tau$-th delayed variable for neuron $i$, taking values $\{S_{ti}^{\tau}:1\leq t\leq T-1\}$ non-sequentially, would mean that the following time point of neuron $i$ depends on previous time point of neuron $i$ in sequence. Dependency of the $\tau+1$-th delayed variable for neuron $i$, taking values $\{S_{ti}^{\tau+1}:1\leq t\leq T-1\}$ non-sequentially, on the $\tau$-th delayed variable for neuron $j$, taking values $\{S_{tj}^{\tau}:1\leq t\leq T-1\}$ non-sequentially, would mean that the following time point of neuron $i$ depends on previous time point of neuron $i$ in sequence. This formulation brings the temporal dependence in sequential neural time series down to dependence between non-sequential variables. This transformation helps to capture the connectivity which is inherently sequential, based on non-sequential variables with algorithms usable for non-sequential data.

The consideration of delayed time series increases the dimensionality, and calls for a robust output never-the-less. Furthermore, in order to maintain stationarity of the data, smaller windows are advantageous. For these purposes, we propose to employ a windowed bootstrap procedure which will enhance the robustness of the inferred connectivity between the signals. The procedure defines time-windows in the time-delayed series over the neurons, $S_{ti}^\tau$, within the total duration sampled. By drawing a time-window of fixed size from any possible overlapping time window on the series, uniformly with replacement, we obtain bootstrapped samples of the series. For example, time-windows of width 100 msecs could be obtained as follows: First draw a uniform random integer $0\leq a\leq T-1$ with replacement  time points to give the window $a_1$ to $a_1+100$ msecs. Second, draw another uniform random integer $0\leq a_2\leq T-1$ with replacement  time points to give the window $a_2$ to $a_2+100$ msecs, and so on. 

The bootstrap step in Neuro-PC is thereby such that for each time-window $t\in W_b$, say of width $\delta_W$, the recordings in the time delayed series $S_{ti}^\tau$ are fetched into the PC algorithm which generates a candidate connectivity graph, say $F_d^{W_b}$. To find the weights of a directed connection in $F_d^{W_b}$ from node i to j, labeled as $M_d^{W_b}$, we perform a linear regression with the time series of node j as response and the time series of node i as covariate. The coefficient to node i found in this regression is considered to be the weight of the directed connection from nodes i to j. We repeat the process for different time windows and obtain the FC graph and matrix for each window.

From the previous step we obtained a collection of connectivity graphs $F_d^{W_b}$ and matrices $M_d^{W_b}$, for each time-window $W_b$. We now consolidate the time-delayed collection into a single time-windowed connectivity graph and matrix between the neurons in the following way. If there is an edge directed from one of neuron $k$'s time delayed series at lower delay to one of neuron $l$'s time delayed series at higher delay, then we draw a directed edge starting from neuron $k$ to neuron $l$. In the connectivity matrix we find the weight of a particular directed connection $i\rightarrow j$ among the neurons by averaging the weights of the connections in the time-delayed series if that connection makes a lower lag of $i$ to be an ancestor of a higher lag of $j$neurons. This is according to the intuition that when we term neuron $i$ has causal influence on neuron $j$, we really mean that the previous time point of neuron $i$ has causal influence on a next time point of neuron $j$. 
For each of the time-windows, $W_{b}$, in the bootstrap procedure we obtain the pair : FC graph $F_d^{W_b}$, and FC matrix $M_d^{W_b}$. To consolidate these time windowed FC components into a single FC graph and matrix, we perform the next step in Neuro-PC, which is the consensus step. The consensus step forms a directed graph $G$ by only having those connections that are present in $50\%$ of the connectivity graphs obtained from the bootstrap iterations. We perform a consensus of the bootstrap connectivity matrices by averaging weights of those connections which are preserved in the resultant connectivity graph. 

\begin{figure}[t]
    \centering
    \includegraphics[width =\textwidth]{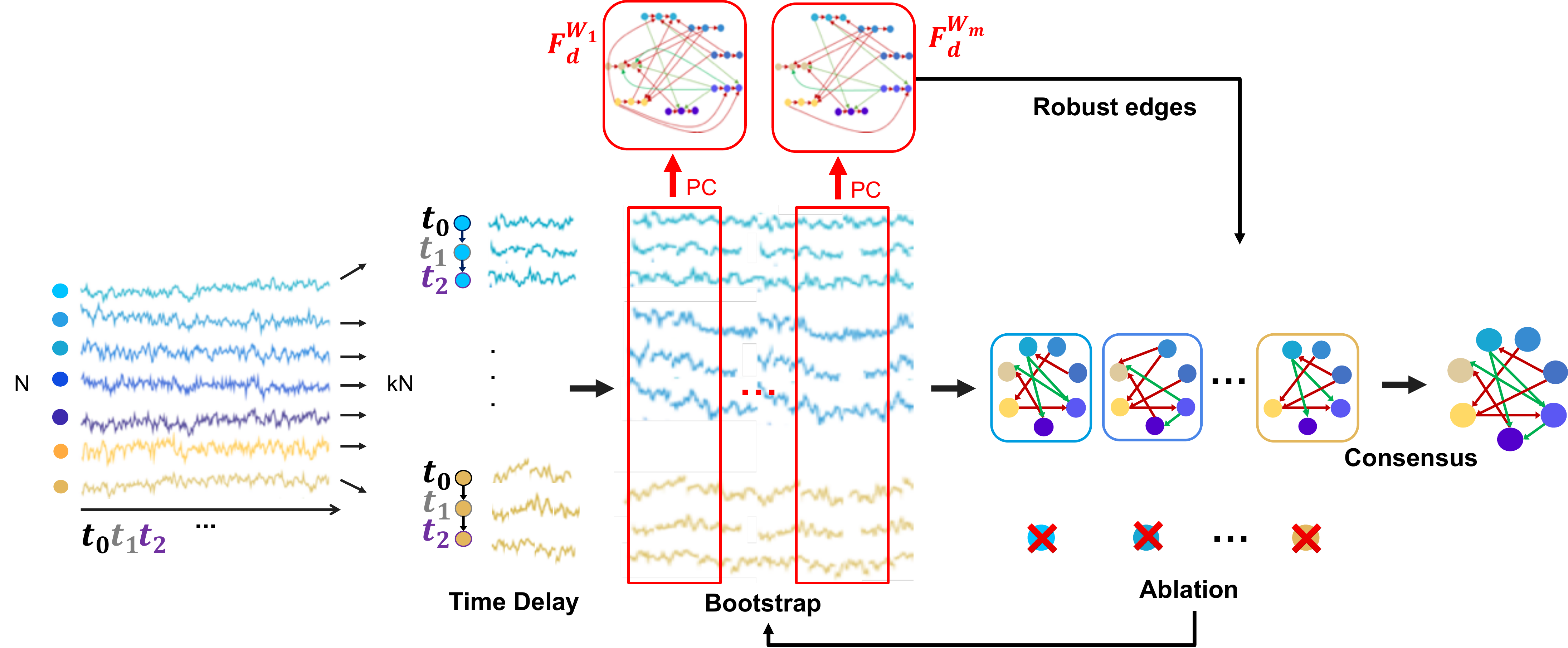}
    \caption{Illustration of the  Time Delay, Bootstrapped PC, Neuron Ablation and Consensus steps of the Neuro-PC algorithm.}
    \label{fig:neuropc}
\end{figure}
We include a final post-processing step to capture any connections that could be missed out by the procedure so far. In this step, we ablate one neuron's time series from the dataset, and proceed with the dataset with one neuron left out. We then repeat the time-delayed series and bootstrapped PC steps upon the new leave-one-out neural time series. This results in a connectivity graph and matrix from each neuron's ablation. We form consensus of these connectivity graphs and matrices to output a single graph and matrix. For consensus, we check for the following conditions to hold: i) $m\rightarrow n$ is present in the connectivity graph obtained when neuron $z$ is ablated, the graph being denoted by $G^{-z}$, ii) $m$ ancestor of $z$ in the connectivity graph $G$ obtained with no neurons ablated, iii) $k\rightarrow n$ is absent in $G$ $\forall k \not\in \{z,n\}$.
If these three conditions are satisfied, then we must have $z\rightarrow n$, because, if $z\not\rightarrow n$, then with the above conditions (i-iii), we have that $m$ is an ancestor of $z$ and $z\not\rightarrow n \forall z\neq n$ in $G$, which implies that, in $G^{-z}$, $m$ is not ancestor of $n$ which again implies that $m\not\rightarrow n$ in $G^{-z}$, contradicting condition (i). So, under conditions (i)-(iii), it must follow that $z\rightarrow n$. With this understanding, in this post-processing step we check if conditions (i-iii) are satisfied. Then if the algorithm did not infer $z\rightarrow n$ in the connectivity graph, in that case, we update the connectivity graph between the neurons by adding the connection $z\rightarrow n$. 
We update the connectivity matrix by including the weights for the added edges by the linear regression approach discussed earlier. We refer to the procedure discussed in this section with the three steps of time delay, bootstrapped PC and ablation as the Neuro-PC algorithm. Figure \ref{fig:neuropc} visualizes the steps of the Neuro-PC. The steps are outlined in Algorithm \ref{algo:neuropc}.

\begin{algorithm}[ht]
\caption{Neuro-PC\label{algo:neuropc}}
\SetAlgoLined
\KwData{$\mathbf{S}$, which is neural recording matrix, of dimension  $N\times T$, for neurons $1, \ldots,N$, and $T$ number of recording time points.}
\KwResult{Causal Functional Connectivity between the neurons}
\Begin{

\textbf{Step 1. Time-Delayed Signals}

Initialize $k$ (default to $5$), maximum time $\tau$\;

Define $S^{(\tau)}_{it} = \mathbf{S}_{kt+\tau,i}$ for $1\leq i\leq N, 1\leq \tau \leq k, 0\leq t \leq T$.
 
Define $\tilde{\mathbf{S}}$, $Nk\times \floor*{\frac{T-1}{k}}$ matrix with elements $S^{(\tau)}_{it}$, columns indexed by $t$, rows by $(i,\tau)$.


 \textbf{Step 2. Bootstrapped PC}

Initialize $n_{iter}$ (default to 50), $t_w$ (default to 500)\;
Assuming, $T>t_w$, else, adjust $t_w$\;
 \For{$iter\leftarrow 1$ \KwTo $n_{iter}$}{
  Sample a column of $\tilde{S}$ uniformly with replacement as starting of a time window, $B_{iter}$ of fixed width $\delta_w$\;
  
  Find $\tilde{G}^{iter}$ = graph output of PC algorithm on $B_{iter}$, with nodes labeled $(i,\tau)$.
  
  
  Find $\tilde{W}^{iter}$:  
  $\tilde{W}^{iter}_{(i,\tau1),(j,\tau2)}=0$ if $(i,\tau1)$ is not an ancestor of $(j,\tau2)$ in $\tilde{G}^{iter}$, else,\\
  $\tilde{W}^{iter}_{(i,\tau1),(j,\tau2)} = $ linear regression coefficient of regressing column index $(i,\tau1)$ on column index $(j,\tau2)$ in $\tilde{\mathbf{S}}$.

\textbf{Step 3. Transforming from Time-Delayed Signals to Neurons}

Define connectivity graph $G^{iter}$ and weight matrix $W^{iter}$ such that 

$i\rightarrow j \in G^{iter} ~ iff ~ (i,\tau1)\rightarrow (j,\tau2) \in \tilde{G}^{iter}$, 

$W^{iter}_{ij} =$ average of $\tilde{W}^{iter}_{(i,\tau1),(j,\tau2)}$ whenever $(i,\tau1)$ is an ancestor of $(j,\tau2)$ in $\tilde{G}^{iter}$,

for some $\tau1<\tau2$, for $1\leq i,j\leq N$.
}

\textbf{Step 4. Robust edges}

 Robust connectivity graph $G$ : $i\rightarrow j \in G$ if $i\rightarrow j \in G^{iter}$ for at least half many $iter$.

Robust connectivity weight matrix $W$: If $i$ not an ancestor of $j$, $W_{ij}=0$, else, $W_{ij} = \frac{\sum_{iter: W^{iter}_{ij}\neq 0}W^{iter}_{ij}}{\sum_{iter} 1\left\{W^{iter}_{ij}\neq 0\right\}}$.
 
\textbf{Step 5. Neuron Ablation and Consensus}

Remove column $i$ from $\mathbf{S}$ and repeat Steps 1-4 to output $G^{(-i)}, W^{(-i)}$, for $1\leq i \leq N$;\ 



Update $G$: If $m\rightarrow n \in G^{(-z)}$, in $G$, $k\not\rightarrow n \forall k\not\in \{z,n\}$, and $m$ is an ancestor of $z$ in $G$, then add to $G$: $z\rightarrow n$,

Update $W$: $W_{zn} = \frac{\sum_{iter: W^{iter}_{zn}\neq 0}W^{iter}_{zn}}{\sum_{iter} 1\left\{W^{iter}_{zn}\neq 0\right\}}$,
for $1\leq z, m, n \leq N$, distinct integers.}
\end{algorithm}



\section{Application to Simulated Data}

We simulate neural dynamics by Continuous Time Recurrent Neural Networks, given in Equation (\ref{ctrnn}), where $u_j(t)$ is the instantaneous firing rate at time $t$ for a post-synaptic neuron $j$, $w_{ij}$ is the linear coefficient to pre-synaptic neuron $i$'s input on the post-synaptic neuron $j$, $I_j(t)$ is the input current on neuron $j$ at time $t$, $\tau_j$ is the time constant of the post-synaptic neuron $j$, with $i,j$ being indices for neurons with $m$ being the total number of neurons.

\begin{equation}\label{ctrnn}
    \tau_j \frac{du_j(t)}{dt}=-u_j (t) + \sum_{i=1}^m w_{ij} \sigma (u_i (t)) + I_j (t), j=1,\ldots, m
\end{equation}

In the simulation study, we consider six motifs ranging over different number of neurons and their connectivity motifs: (1) $4$ neurons with $w_{13}=w_{23}=w_{34}=50$; (2) $4$ neurons with  $w_{13}=w_{23}=w_{43}=40$, (3) $5$ neurons with $w_{13}=w_{24}=w_{35}=w_{45}=80$, (4) $7$ neurons with $w_{21}=w_{31}=w_{41}=w_{51}=w_{61}=w_{71}=20$, (5) $10$ neurons with $w_{21}=w_{31}=\ldots=w_{10,1}=20$, (6) $13$ neurons with $w_{21}=w_{31}=\ldots=w_{13,1}=20$. The time constant $\tau_i$ is set to 10 msecs for each neuron $i$. We consider $I_i(t)$ to be distributed as independent Gaussian process with the mean of 2 and the standard deviation of 2. The signals are sampled at a time gap of $e \approx 2.72$ msecs for a total duration of $1000$ msecs. 

\begin{figure}
    \centering
    \includegraphics[width = 5 in]{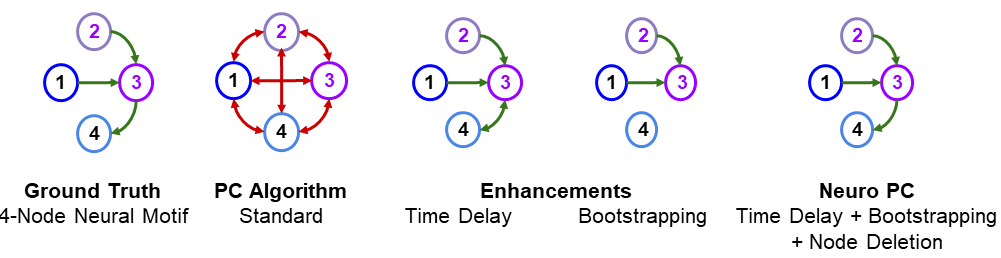}
    \caption{Inference of the FC graph (first degree connections) from the firing rate series simulated by the Continuous Time Recurrent Neural Network on the motif outlined as Ground Truth. Left to right: Ground Truth compared with the outcome of PC algorithm, variants of its enhancements: PC algorithm+Time Delay,  PC algorithm+ Bootstraping, NeuroPC: PC algorithm+Time Delay+Bootstraping+Node Deletion.}
    \label{fig:fig4}
\end{figure}

We compute the inferred connectivity graphs from the simulated data by the PC algorithm and by Neuro-PC and compare them with the Ground Truth connectivity motifs set for the network. We show an example in Fig.~\ref{fig:fig4} that demonstrates the computed graph by the PC Algorithm and the contribution of each step in Neuro-PC to refine the graph. In particular, this examples demonstrates the typical performance of the PC Algorithm which provides multiple edges which should be filtered out. The time delay step of the Neuro-PC algorithm ensures that the edges, that are essentially inter-temporal, are incorporated with greater accuracy and spurious edges are excluded. The bootstrapping step ensures more robustness of the inferred connectivity by excluding those edges which appeared in less than $50\%$ of the graphs inferred from the bootstrap iterations. 

\begin{figure}[t!]
    \centering
    \includegraphics[width = \textwidth]{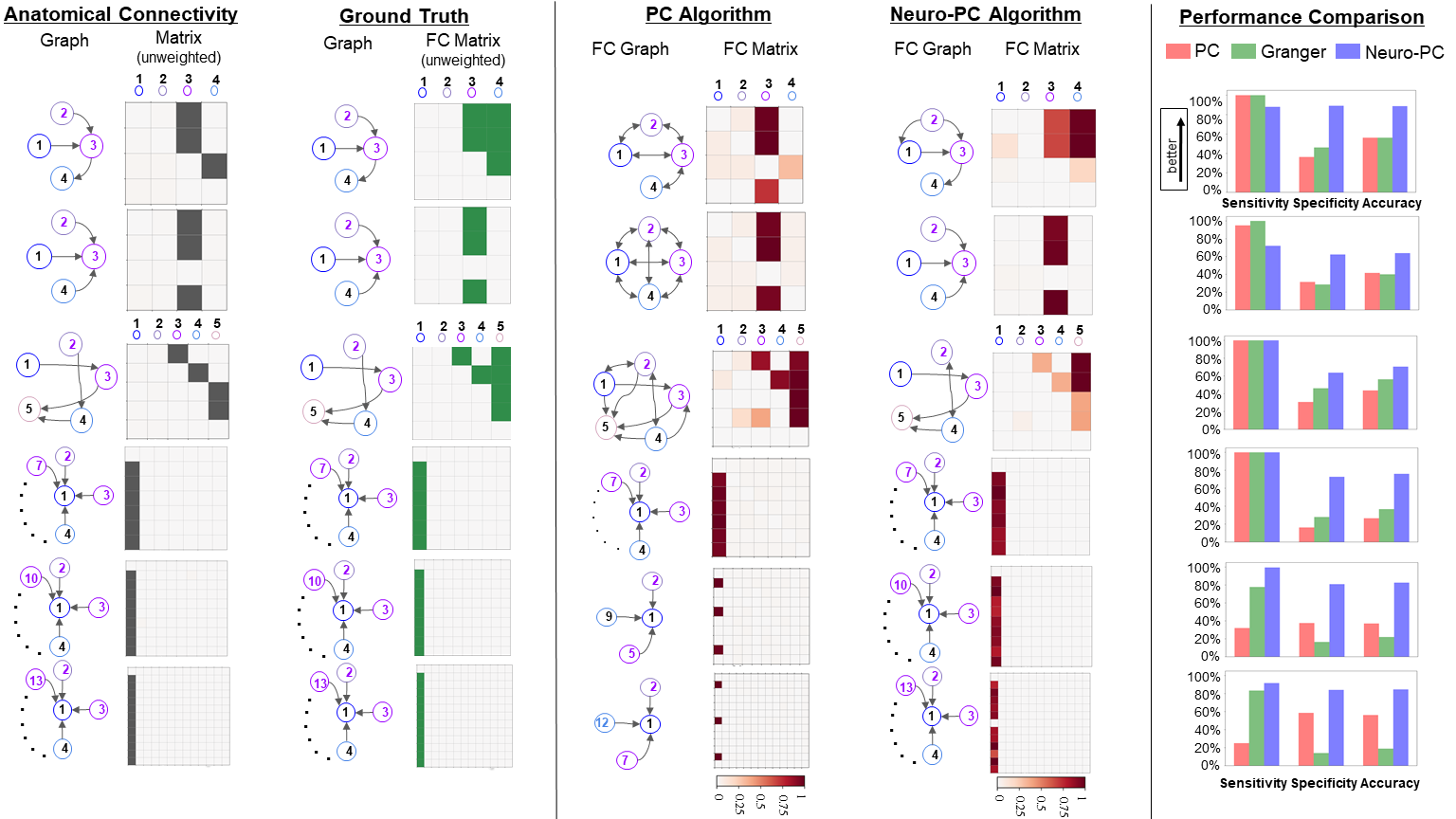}
    \caption{The performance of PC algorithm and Neuro-PC algorithm in inference of ancestral graph (FC) denoted as Ground Truth from neural firing rates for six different motifs, ranging over four to thirteen neurons. The ground truth FC matrix is devoid of weights, so are colored green. The results of the algorithms output the FC graph and the FC weights which are colored according to the inferred weights. We observe that the Neuro-PC recovers the Ground Truth with greater accuracy, specificity and comparable to greater sensitivity than PC algorithm.
}
    \label{fig:fig5}
\end{figure}

Figure \ref{fig:fig5} displays the causal FC graphs and matrices inferred from the six simulations using Neuro-PC as well as of the PC algorithm. Each of the two algorithms output a connectivity graph and matrix. In the graph, $i\rightarrow j$ means that there is a functional connection from neuron $i$ to neuron $j$. The FC matrix provides weights to direct as well as indirect connections, where in, entry $(i,j)$ has a non-zero weight if there is a path from neuron $i$ to neuron $j$ in the FC graph. The Ground Truth FC of the motifs is based on the weights $w_{ij}$ in equation (\ref{ctrnn}), where, $i\rightarrow j$ in the Ground Truth FC graph if $w_{ij}\neq 0$, and entry $(i,j)$ of the Ground Truth FC matrix has a non-zero weight if there is a path from $i$ to $j$ in the Ground Truth FC graph. 

We quantified the performance of the algorithms by first estimating FC matrices for 50 simulated samples with the same specifications from each motif as discussed in Section 5. The sensitivity or true positive rate is calculated by the proportion out of the non-zero entries of the estimated FC matrices that were truly non-zero. Specificity or true negative rate is calculated by the proportion out of the zero entries of the estimated FC matrices that were truly zero. Accuracy is calculated by the proportion of the entries of the estimated FC matrices that matched with the corresponding entry in the ground truth FC matrix. Neuro-PC recovers the structure of the Ground Truth with greater specificity and accuracy exceeding by more than $20\%$ for all the six motifs and at least as good sensitivity for Motifs in rows 3-6 in Figure \ref{fig:fig5}.

We also compared the performance with Granger Causality for all the six motifs. Granger Causality is commonly implemented in terms of multivariate auto-regressive model in multivariate time series scenario \citep{geweke1982measurement}. In the implementation, it is tested whether the addition of a prediction of the time-series from another time-series through a multivariate auto-regressive (MVAR) model may improve our prediction of the present behavior of the time-series \citep{ding200617}. We use the \emph{Nitime} Python library \citep{rokem2009nitime} for the analysis to fit a MVAR model of order 3 with sampling interval of $e\approx 2.72$ msecs, followed by computation of the Granger Causality by the \emph{GrangerAnalyzer} function. The specificity, sensitivity and accuracy chart in Figure \ref{fig:fig5} shows that Neuro-PC performed comparatively better in recovering the ground truth  with greater than 20\% improvement in specificity and accuracy in Motifs 1-2, 4-6 in Figure \ref{fig:fig5}.

\section{Application to Neurobiological Data}
 We use the Visual Coding Neuropixels database from the Allen Brain Observatory ~\citep{de2020large,allenbrainobs}. The database consists of sorted spike trains and local field potentials recorded simultaneously from up to six cortical visual areas, hippocampus, thalamus, and other adjacent structures of mice, while the mice passively view a stimuli shown to them. The stimuli include static gratings, drifting gratings, natural scenes/images and natural movies, which are shown to the mice with repetitions. The data has been recorded from the neurons with the recently developed technology of Neuropixels which allows real-time recording from hundreds of neurons across the brain simultaneously by inserting multiple probes into the brain~\citep{neuropixels}.

For the purpose of application and comparison of the results of the methods discussed in this paper, we restrict our analysis to a 116 days old male mouse (Session ID 791319847) with 555 neurons whose spike trains are recorded simultaneously by six Neuropixel probes. The spike trains during the entire experiment were recorded at a frequency of $1$ KHz.  
We analyze the spike trains for four stimuli: 
\begin{enumerate}[leftmargin=*]
    \item Natural scenes, consisting of 118 natural scenes selected from three databases (Berkeley Segmentation Dataset, van Hateren Natural Image Dataset and McGill Calibrated Colour Image Database), with each scene presented briefly for 250ms and then replaced with the next scene image. Each scene is repeated 50 times in random order with intermittent blank intervals.
    \item Static gratings consisting of full-field sinusoidal gratings with 6 orientations (the angle of the grating), 5 spatial frequencies (the width of the grating), and 4 phases (the position of the grating) resulting in 120 stimulus conditions. Each grating is presented briefly (250 ms) before being replaced with a different orientation, spatial frequency and phase condition and each condition is repeated 50 times, in random order with intermittent blank intervals. 
    \item Gabor patches with 3 orientations where the patch center is lying at one of the points in a $9\times9$ visual field. Each gabor patch is being presented for 250ms and then replaced by a different patch, and each condition is repeated 50 times in random order with intermittent blank intervals.
    \item Full-field flashes, lasting for 250 ms followed by a blank interval of 1.75 s, and then the next flash, totaling 150 repetitions.
\end{enumerate}
This variety of stimuli is ranging from relatively \textit{natural stimuli} invoking mice's natural habitats (natural scenes) to \textit{artificial stimuli} (static gratings, gabors and flashes). Among the artificial stimuli, static gratings incorporate sinusoidal patches, while full-field flashes incorporate sharp changes in luminosity in the whole visual field in short period of time, and gabor patches incorporate sinusoidal patches with declining luminosity with distance from the center of the patch. With this choice of four stimuli we investigate how the variety of stimuli possibly invokes distinct patterns of neuronal interactions and connectivity. We exclude dynamic stimuli like natural movies, and drifting gratings, from this analysis because their results would require more nuanced study and interpretation, which we defer for future analysis.

\begin{figure}[t]
    \centering
    \includegraphics[width= 5in]{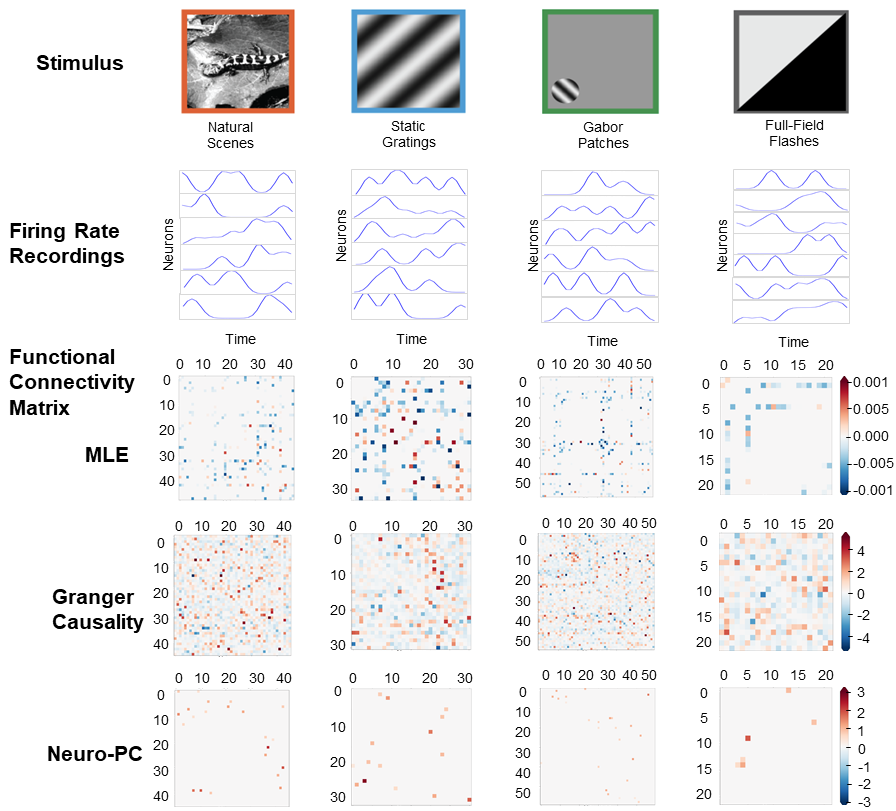}
    \caption{Comparison and demonstration of the FC inferred for a benchmark of mice brain data obtained from the Allen Institute’s Neuropixels dataset, by three methods for FC inference: FC matrix using Undirected FC estimated by Graphical LASSO penalized MLE, Granger Causality and Neuro-PC. The colors in the FC matrix have been normalized within each method by the minimum and maximum of values outputted by the method.
    In the experiment, the mice were subject to different stimuli. For the benchmark we selected four such stimuli with different characteristics: Natural Scenes, Static Gratings, Gabor Patches and Full-Field Flashes. 
}
    \label{fig:fig6}
\end{figure}
\begin{table}[b]
    \centering
    \begin{tabular}{l|l|l|l}
    \hline
    Method & Noise-level & Outcome graph & Causality\\
    \hline
    \hline
    MLE& Noisy & Undirected & Non-causal\\
    Granger Causality & Noisy &  Directed & Concerns on causality\\
    Neuro-PC & Less noisy & Directed & Markov Property for causality\\
    \hline
    \end{tabular}
    \caption{Comparison of results of the three methods applied in the dataset}
    \label{tab:comparison_on_data}
\end{table}

\textbf{Data Processing:} We convert the spike trains recorded at 1 KHz to bin size of 10 ms by aggregating and then separating by start and end times of each stimuli presentation and obtain the Peri-Stimulus Time Histograms (PSTH) with bin-size 10 ms. We smooth the PSTHs by a Gaussian smoothing kernel of bandwidth $16$ms which provides a smoothed version of the PSTH for each neuron and each stimulus presentation. Some examples of the smoothed PSTH are displayed in Figure~\ref{fig:fig6}. We use the smoothed PSTHs for neurons over each stimuli type as input for inference of the FC between the neurons for each stimuli presentation. For each stimulus presentation, we first selected the set of neurons that were active in at least $25\%$ of the bins in the PSTH. This was done since the inference of functional connections between the inactive neurons is expected to be noisy and unreliable. In Neuro-PC inference, to further cope with noisy estimates, we excluded the connections that were present in less than $50\%$ of the bootstrap iterations, and those connections between pairs of neurons which were detected with p-value of more than $0.4$ in the skeletal graph outputted by the PC algorithm. 

\textbf{Results:} We used three methods for inferring the FC from neural signals and compare their results: 1) Undirected FC inferred by Graphical LASSO penalized Maximum Likelihood Estimation (MLE),  2) Granger Causality, and 3) Neuro-PC (Our). A summary of the results is provided in Figure~\ref{fig:fig6}. Table~\ref{tab:comparison_on_data} outlines comparisons of the results from the three methods on three key aspects - noise directionality and causal interpretation. 

Visual inspection indicates that gabor patches and full-field flashes result in a sparser matrix in comparison to natural scenes and static gratings for each of the three methods. This can be intuitively understood since natural scenes mimic the mouse's natural habitat and static gratings have a pattern covering the entire visual field of the mouse. These stimuli could encourage a stronger and more intricate response in comparison to gabor patches, which occupy a small fraction of the visual field and flashes which even though are full-field, are characterized by only a heightened luminosity and devoid of patterns like gratings or rich content. 

Examining the Neuro-PC graph structure, we observe that natural scenes and gabor patches lead to a more 'asymmetric connectivity' graph in contrast to static gratings and flashes. This global structure could reflect the specific features of the stimuli. Both natural scenes and gabor patches have features that encourage attention asymmetrically on the visual field, unlike static gratings and full-field flashes which are symmetric over the visual field. In comparison, the MLE always outputs a symmetric connectivity graph, and Granger Causality doesn't reflect a distinct asymmetry.

\begin{figure}[t!]
    \centering
    \includegraphics[width= \textwidth]{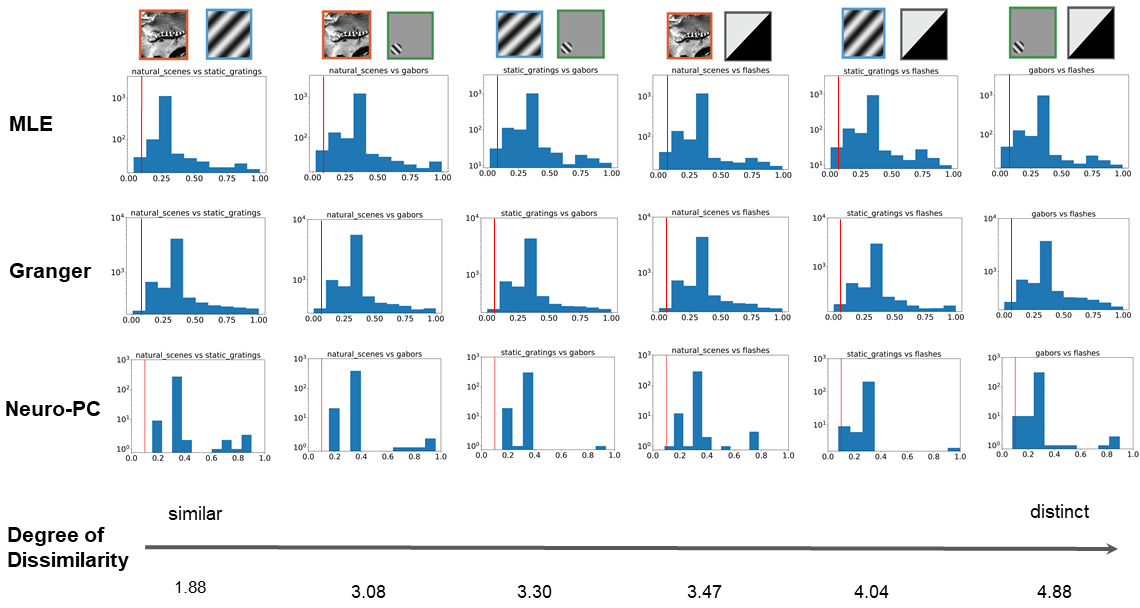}
    \caption{This figure outlines the results of hypothesis test for changing FC between pairs of neurons over change in stimuli.}
    \label{fig:fig7}
\end{figure}

We further analyze whether change in stimulus type alters significantly the functional connectivity between any pair of neurons. In particular, we performed statistical tests to determine whether each cell in the FC matrix has a value that remains the same between pairs of stimuli-types. We use a paired two-sample t-test for the purpose. We considered 30 trials for each stimulus-type that generated 30 trials for each cell of the connectivity matrix. For each pair of stimulus-type, we plotted a histogram of the p-values, where each p-value corresponds to testing whether a cell of the FC matrix incurred a change between the stimulus pair. The results are shown in Figure \ref{fig:fig7}.

We observe from the histograms in Figure \ref{fig:fig7} that both MLE and Granger Causality do not exhibit visible changes in the distribution of the p-values for different stimuli. This is indicative of similar comparisons associated with distinct pairs of stimulus-type. Whereas in the case of Neuro-PC, the p-value distributions reflected visible changes for different stimulus-type pairs. The more sparse result from Neuro-PC compared to the other approaches helps in noting changes in the histograms by visual inspection. 
Indeed, functional connections are expected to be sensitive to stimuli~\citep{stimu_fc_cats,stimu_fc_mouse1}, such that for different stimulus pairs, a change in the distribution of the p-values is warranted. Furthermore, Neuro-PC shows that some functional connections changed significantly (the connections whose p-value lie to the left of the red 0.05 p-value line) as stimulus changed between natural scenes and flashes, static gratings and flashes and gabor patches and flashes, which gives insight into those pairs of neurons with functional connections that are responsive to the stimulus change. The neurons which showed significant change of response with change in stimulus would be of interest in studying neurobiological response to these stimuli.

For each pair of stimuli, we can also quantify the degree of functional dissimilarity between the pair of stimuli by the percentage of connections which showed strong change with switch between the two stimuli (having p-value less than 0.3) given by Neuro-PC. Such analysis indicates that natural scenes vs static gratings were most similar with respect to this metric, while, least similar were natural scenes vs flashes, static gratings vs flashes and gabors vs flashes. This can be intuitively understood since natural scenes and static gratings both provide rich content and both cover the full visual field, whereas, flashes lack rich content invoking less neuronal interactions. Gabor patterns may evoke distinctly different neuronal interactions than flashes since the former does not cover the entire visual field while the latter does. 

\section{Discussion}
In this paper, we propose a novel methodology for inference of causal functional connectivity between neurons from multiple neural time series. We demonstrated the accuracy and robustness of the method in simulations and performed a comparative study with other related methods in the literature. We applied the method to a neurobiological dataset of signals from the brain of the mouse and compared with other approaches. The results give insights into the FC between the neurons in the mouse brain in a variety of stimuli scenario, for example, how the FC varies with stimuli, whether changes in stimuli are able to invoke a significant change in the FC structure, and subsequently, similarity in stimuli in terms of joint connectivity structures.


Notably, while the importance and the need of causal FC from neuronal recordings is recognized in literature, e.g., as described Reid et al.~\citep{reid2019advancing} there is currently a gap in the literature in obtaining the causal FC from neuronal recordings~\citep{caus_fc_1}. Existing methods for inference of FC obtain the associative and undirected connectivity with statistical guarantees~\citep{assoc_fc_1,assoc_fc_2}. The focus of the current paper is to contribute to finding the causal FC from multiple neuronal time series based on directed Probabilistic Graphical Models (PGM)\citep{koller2009probabilistic}. It provides a formulation which shown effective in finding causal networks in other disciplines \citep{wang2017potential,sinoquet2014probabilistic, mourad2012probabilistic, wang2005new, friedman2004inferring}. Directed PGM provides an alternative approach to causality than Granger Causality which have been used in neuroscience literature to characterize causality. However, Granger Causality has difficulties in the presence of common causes \citep{mehler2018lure, maziarz2015review} and it depends on the MVAR autoregressive linear model assumption in it's popular implementation \citep{geweke1982measurement}. The PGM, on the other hand, deals with common causes is a non-parametric approach not requiring parametric model assumptions. 
The key step in the Neuro-PC algorithm is obtaining the directed PGM satisfying the Markov Property, between the neuronal time series \citep{spirtes2000causation}. 
The Markov property is the principle that if we fix the states of neurons that directly influence a neuron of interest, $X$, then the states of neurons that are connected indirectly to $X$ through direct connections are rendered independent with respect to $X$ \citep{glymour2019review}. With regards to unobserved confounding variables, in the basic setup,  Neuro-PC algorithm does not account for them, similarly to Granger Causality. However, by switching the PC algorithm with the FCI algorithm \citep{spirtes2000causation} in Step 2 (Bootstrap) of Neuro-PC, could account for unobserved confounding variables, because the FCI algorithm is a modification of the PC algorithm that takes into account unobserved confounding variables.

In addition to having the PC algorithm at its core, Neuro-PC includes components targeted towards the particular neural interactions. In neuroscience literature, causality is often defined as a cause of an observed neural event (the ‘effect’) as a preceding neural event whose occurrence is necessary to observe the effect~\citep{reid2019advancing}. The approach of time delay incorporates this definition into Neuro-PC since it essentially identifies whether the previous time values of the neurons impact the present value of a particular neuron by the obtained directed graph. Bootstrapping step gets rid of spurious connections by repeating the inference of causal FC over several subsampled blocks of the neuronal time series and discarding those connections that are absent in half the repetitions. The neuron ablation step is similar to the practice of interrogation of neural circuits \citep{emiliani2015all} and intervention in statistical causal inference literature \citep{schulz2007learning}. 

The Neuro-PC algorithm is applicable to a wide range of recordings. In principle, it can be applied to any continuous neural signal,
such as spiking rates of individual neurons, calcium dynamics of individual neurons, fMRI recordings, and also across organisms and accross parts of the nervous system. We have investigated in the current paper the FC from the brain of a 116 days old male mouse for application of the Neuro-PC and comparison with related approaches. A noteworthy aspect of the dataset used in this paper is that it consists of simultaneously recorded spike-trains from neurons by probes inserted within the mouse brain \citep{steinmetz2018challenges_neuropixels,allenbrainobs}. While existing literature of FC focused primarily on fMRI recordings from voxels in the brain that represent aggregate states of thousands of neurons~\citep{fmri_van2010exploring, caus_fc_fmri}, the advent of Neuropixels opens the doors to individual neuron recordings. 

In further research, we would perform a more detailed analysis of the FC of the mouse brain with more insights into the neurons involved in FC in terms of their locations in the brain and comparisons of the neurons in FC over different stimuli, multiple mice, and mice of different age and gender. We also aim to investigate the causal FC of other organisms such as \emph{C. elegans} and Zebra fish. 

\bibliographystyle{unsrt}
\bibliography{main}

\end{document}